\documentclass{ws-procs9x6}

\usepackage{amssymb}
\usepackage{amsmath}

\newcommand{\benumerate}{\begin{enumerate}}
\newcommand{\eenumerate}{\end{enumerate}}

\newcommand{\bitemize}{\begin{itemize}}

\newcommand{\eitemize}{\end{itemize}}
\newcommand{\der}[2]{\frac{\partial #1}{\partial #2}}

\newcommand{\dersec}[2]{\frac{\partial^{2} #1}{\partial #2^{2}}}
\newcommand{\derthree}[2]{\frac{\partial^{3} #1}{\partial #2^{3}}}

\begin{document}

\title{On the models of nonlocal nonlinear optics.\footnote{\uppercase{T}his
 work is partially supported by \uppercase{COFIN PRIN SINTESI}
2004.}}

\author{Boris G. Konopelchenko and Antonio Moro}

\address{Dipartimento di Fisica dell'Universit\`a di Lecce \\
and Istituto Nazionale di Fisica Nucleare, Sezione di Lecce \\
via Arnesano,  I-73100, Lecce \\
E-mail: konopel@le.infn.it, antonio.moro@le.infn.it}

\maketitle

\abstracts{We show that under certain assumptions a general model
of nonlocal nonlinear response in $1+1-$dimension is equivalent to
the model considered by Kr\'olikowski and Bang for a Kerr-type
medium. We derive the limit of weak nonlocality in high frequency
regime and discuss the integrable cases.}

\section{Introduction}
A paraxial laser beam propagating in a medium with response of the
form ${\bf D} = \varepsilon_{0} {\bf E} + \sigma^{3} {\bf
D}^{(3)}$ ($\sigma <<1$) is described by the following equation
\begin{equation}
\label{wave_eqt} 2 i \sqrt{\varepsilon_{0}} \omega \der{{\bf
E}}{z} + \nabla^{2}_{\bot} {\bf E} + \omega^{2} {\bf D}^{(3)} = 0,
\end{equation}
where ${\bf D}^{(3)}$ is a certain function of the electric field.
For derivation of such type of equation
see~[\refcite{Landau,Sulem}]. Recent theoretical and experimental
results highlight a very interesting phenomenology, such as
``accessible solitons", collapse arrest, optical vortices (see
e.g.~[\refcite{Snyder}$-$\refcite{Krolik2}] and references
therein), associated with so-called nonlocal nonlinear media. Most
of the models studied before are not integrable and exact
analytical solutions are not available for them. However,
recently, it was shown that certain choice of nonlocal responses
leads to $2+1$-dimensional integrable models~\cite{Moro1,Moro2}.

In the present paper, for sake of simplicity we discuss the
$1+1-$dimensional case. In the section~\ref{sec_Schroedinger}, we
consider a more fundamental nonlocal nonlinear model than one
considered in the paper~[\refcite{Moro1}]. We specify under  which
conditions it leads us to a generalization of the model discussed
by Kr\'olikowski and Bang~\cite{Krolik3}. The model proposed
in~[\refcite{Krolik3}] is of interest since it can be solved
analytically for weak nonlocality and it admits soliton-like
solutions.

In the section~\ref{sec_high_freq} we analyze a class of weak
nonlocal responses in the high frequency regime. The study of high
frequency limit can help to detect interesting properties of the
beam such as singular phases and consequent vortex type
behaviours. Moreover, this study could be useful to construct new
nontrivial ansatz for applying a variational method. Looking for
solutions which depend ``slowly" on the coordinate along the
propagating direction, we show that the phase of the electric
field is given by an over-determined system of partial
differential equations for the phase. Their compatibility and some
integrable cases are also discussed.

\section{The nonlocal nonlinear Schr\"odinger equation.}
\label{sec_Schroedinger}
Equation~(\ref{wave_eqt}) in
$1+1-$dimensional case looks like as follows
\begin{equation}
\label{1Dwave_eqt} 2 i \omega \der{{\bf E}}{z} + \dersec{\bf E}{x}
+ \omega^{2} {\bf D}^{(3)} = 0.
\end{equation}
We set $\varepsilon_{0} = 1$ without lost of generality. Let us
assume a constitutive relation of the following general form
\begin{equation}
\label{constitutive} {\bf D}^{(3)} = \int_{-\infty}^{+\infty}
R(x-x';a) \; N \left(I(x') \right ) \; {\bf E}(x')  dx'.
\end{equation}
Distribution $R(x-x';a)$ characterizes the nonlocal response
around the point $x$ and $a$ is the ``width" parameter (in the
following it will be assumed to be depending on the frequency
$\omega$). $N(I)$ is an arbitrary nonlinear response depending on
the intensity of the electric field $I = \left |{\bf E}
\right|^{2}$. In what follows we will propose a way to simplify
the general model~(\ref{constitutive}) which will lead us to model
considered in the paper~[\refcite{Krolik3}].

Let us consider a nonlocal distribution $R\left (x-x';a \right)$
of width $\delta R$ defined as the minimum such that
\begin{equation}
R \left (x-x';a \right) \simeq 0, \quad{} \forall x' \notin \left
[ x- \delta R, x+ \delta R\right ].
\end{equation}
Analogously we introduce the widths $\delta E$ and $\delta N$ of
the electric field and the nonlinear response respectively.
Suppose they verify the following conditions
\begin{equation}
\label{wideness} \delta R \sim \delta N, \quad{} \delta R <<
\delta E.
\end{equation}
Let us assume a nonlinear response of the form
\begin{equation}
\label{N_gamma} N \left (I (x) \right) = \tilde{N} \left (X
\right),
\end{equation}
where $X = \gamma \: x$ and $\gamma =  1/ \delta N$. Expanding
${\bf E}(x')$ and $N\left (I(x') \right )$ in Taylor series around
$x$, one gets the following approximation of the
formula~(\ref{constitutive})
\begin{equation}
\label{bang_model} {\bf D}^{(3)} \simeq \left ( \int_{- \infty}^{+
\infty} R \left (x-x';a \right) N \left (I(x') \right) dx' \right
) {\bf E}(x),
\end{equation}
where we kept into account that due to the
equation~(\ref{N_gamma}) higher orders of the expansion of $N\left
(I(x) \right)$ are not negligible. Note that the
formula~(\ref{bang_model}), in the case of nonlocal Kerr-type
medium, leads to the nonlocal nonlinear Schr\"odinger equation
discussed in the paper~[\refcite{Krolik3}].

\noindent For instance, given a bell-shape electric field ${\bf E}
= {\bf E}_{0} \; \exp \left [- x^{2}/2 \sigma^{2} \right]$, a
nonlinear response of the form $N(I) = I^{\alpha} = \left |{\bf
E}_{0} \right |^{2 \alpha} \; \exp \left [- \left (\gamma x
\right)^{2}/{2 \sigma^{2}} \right]$, where $\gamma = \sqrt{2
\alpha}$, satisfies the condition~(\ref{N_gamma}). Nevertheless,
it's easy to see that validity of relations~(\ref{wideness}) is
sufficient to obtain the model~(\ref{bang_model}). For instance,
choosing ${\bf E} = {\bf E}_{0}/ \cosh^{2}(x)$  and $N \left (I
\right ) = I^{\alpha}$, condition~(\ref{wideness}) is verified for
$\alpha$ large enough.

\section{Weak nonlocality and high frequency limit.}
\label{sec_high_freq}

Let us consider a limit of small nonlocality ($\delta R << 1$)
such that it is reasonable to expand both ${\bf E}$ and $N$ in
Taylor series. This expansion gives
\begin{gather}
\label{constitutive_weak}
\begin{aligned}
{\bf  D} =& N \left (I(x) \right) {\bf E}(x) + R_{1} \left (N
\der{{\bf E}}{x} + \der{N}{x} {\bf E} \right) + \\
&R_{2} \left ( \frac{1}{2} N \dersec{{\bf E}}{x} + \der{N}{x}
\der{{\bf E}}{x} + \frac{1}{2} \dersec{N}{x} {\bf E} \right) +  \\
&R_{3} \left(\frac{1}{6} N \derthree{{\bf E}}{x} + \frac{1}{2}
\der{N}{x} \dersec{{\bf E}}{x} + \frac{1}{2} \dersec{N}{x}
\der{{\bf E}}{x} + \frac{1}{6} \derthree{N}{x} {\bf E} \right ) +
\dots
\end{aligned}
\end{gather}
where the distribution $R\left (x-x';a \right)$ is assumed to be
normalized
\begin{equation}
\int_{-\infty}^{+\infty} R \left(x-x';a \right) dx' = 1
\end{equation}
and $R_{n}$ are the $n$-moments
\begin{equation}
R_{n} = \int_{-\infty}^{+\infty} R\left(x-x';a \right) \; \left(
x-x'\right)^{n} \;dx'.
\end{equation}
In the present section, we are interested to perform the high
frequency limit. In particular, we assume that for $\omega$$\to
\infty$ the effective nonlocality decreases because of rapid
oscillation of the electric field. So, one has
\begin{equation}
\lim_{\omega \to \infty} R \left(x-x';a(\omega) \right) = \delta
\left(x-x' \right),
\end{equation}
where the parameter $a$ is a function of the frequency such that
there exists its limit  for $\omega \to \infty$ and $\delta
\left(x-x' \right)$ is the Dirac $\delta$-function.

As illustrative example, we focus on the distribution defined as
follows

\begin{gather}
\label{distribution1} R \left (x-x';a \right ) = \; \left \{
\begin{aligned}
\frac{3}{4} \left (- a^{3} (x-x')^{2}+ a \right) \quad{} x' \in
\left [x-\frac{1}{a}, x + \frac{1}{a} \right ] \\
0 \quad{}\quad{} x' \notin \left [x-\frac{1}{a}, x + \frac{1}{a}
\right ]
\end{aligned}
\right .
\end{gather}
It is straightforward to verify that
distribution~(\ref{distribution1}) tends to the $\delta-$function
as $a \to \infty$, that is
\begin{gather}
\begin{aligned}
\int_{-\infty}^{\infty} R \left (x-x';a \right) \;dx' &= 1 \\
\lim_{a \to \infty} \; \int_{-\infty}^{+\infty} R \left(x-x';a
\right) \; f(x') \; dx' &= f(x).
\end{aligned}
\end{gather}
A direct calculus shows that
\begin{equation}
\label{moment_behavior} R_{2n} = \frac{r_{2n}}{a^{2n}}, \quad{}
R_{2n + 1} = 0, \quad{} n \in \mathbb{N} \cup \left\{0\right\}.
\end{equation}
We would like to stress that the present discussion still holds
for any nonlocal response whose moments are of the
form~(\ref{moment_behavior}).

Let us consider a general dispersion law of the following form
\begin{equation}
\label{Cole_form} N = N_{0} + \frac{\tilde{N}}{1 + \left ( i
\tau_{0} \omega \right )^{2 \nu}},
\end{equation}
where $\tau_{0}$ is usually called relaxation time. In the high
frequency limit $\omega \to \infty$, expanding $I = I_{0} +
\omega^{-2\nu} I_{1} + \dots$,  one gets
\begin{equation}
N = N_{0}\left(I_{0} \right) + \frac{N_{1}\left( I_{0},
I_{1}\right)}{\omega^{2 \nu}} + O \left(\frac{1}{\omega^{4 \nu}}
\right),
\end{equation}
where $N_{1}$ is complex-valued and $2 \nu > 0$. For $2 \nu < 1$,
relation~(\ref{Cole_form}) is known as Cole-Cole dispersion
law\cite{Cole} which substitutes the ``classical'' Debye law ($2
\nu = 1$) for a wide class of liquid and solid polar media.

We perform the high frequency limit in usual way looking for
solutions of the form
\begin{equation}
{\bf E} = {\bf E}_{0} \: e^{i \omega S}.
\end{equation}
Moreover, we assume that the electric field depends slowly on the
$z-$variable according to the rule
\begin{equation}
\label{zrule} \der{}{z} \rightarrow \omega^{-2 \nu} \der{}{z}.
\end{equation}
It is straightforward to see that under the assumption given above
and under the choice $\alpha = 1 + \nu$, the high frequency limit
of equation~(\ref{1Dwave_eqt}) leads, at $\omega^{2}$ and
$\omega^{2- 2\nu}$ orders, to the following system
\begin{gather}
\label{phase_even}
\begin{aligned}
S_{x}^{2} &= N_{0} \\
S_{z} &= - \frac{r_{2}}{4} N_{0} S_{x}^{2} + \frac{1}{2} N_{1},
\end{aligned}
\end{gather}
where $f_{x} = \partial f /\partial x$.

\noindent System~(\ref{phase_even}) is over-determined and its
compatibility condition is equivalent to the following nonlinear
equation
\begin{equation}
\label{N0_even} \pm N_{0}^{-\frac{1}{2}} \; N_{0z} = - r_{2} N_{0}
\; N_{0x} + N_{1x}.
\end{equation}
For $N_{1} \neq 0$ solutions of equation~(\ref{N0_even}) are
complex-valued, and they implies complex-valued phases. Such
solutions are connected in principle with absorption or
amplification of the electric field.

We note that, once $R$ is assigned, the model~(\ref{bang_model}),
at the order $\omega^{2-2\nu}$, provides us with the equation
\begin{equation}
\label{bang_Sz} S_{z} = \frac{1}{2} N_{1},
\end{equation}
and the compatibility condition can be written down in the form of
conservation law $M_{0z} = N_{1x}$ where $M_{0} = \pm
\sqrt{N_{0}}$.

If we assume $N_{1} = 0$, setting $N_{0} = \varphi^{\frac{2}{3}}$
equation~(\ref{N0_even}) is reduced to the well known Burgers-Hopf
equation
\begin{equation}
\label{burgers} \varphi_{z} = \pm \varphi \; \varphi_{x}.
\end{equation}
It is solvable with the hodograph method and the solutions are
given in terms of the following implicit relation
\begin{equation}
\label{burgers_sol} x \pm \varphi \; z + H \left (\varphi \right)
= 0,
\end{equation}
where $H$ is an arbitrary function of its argument.

Now, let us assume $N_{1} = N_{1} \left (N_{0} \right)$. Even in
this case, equation~(\ref{N0_even}) is reduced to the following
quasilinear equation
\begin{equation}
\label{N0_complex} N_{0z} = \pm N_{0}^{\frac{1}{2}} \left (
N_{1}'\left (N_{0} \right) - r_{2} N_{0} \right) N_{0x},
\end{equation}
where $ N_{1}' = d N_{1}/dN_{0}$.

 Analogously to the Burgers-Hopf equation it is solved by the
hodograph relation
\begin{equation}
\label{N0_complex_sol} x \pm N_{0}^{\frac{1}{2}} \left(
N_{1}'\left(N_{0} \right)- r_{2} N_{0} \right) z + L\left (N_{0}
\right) = 0,
\end{equation}
where $L$ is an arbitrary function of $N_{0}$.

Other integrable equations can be obtained considering nonlocal
response of the following form
\begin{gather}
R \left (x-x';a \right) = \left \{
\begin{aligned}
\frac{n+1}{1- (-1)^{n} + 2 n} \left(- a^{n+1} \left(x-x'
\right)^{n} + a \right), \quad{} x' \in D  \\
0, \quad{}x' \notin D
\end{aligned}
\right .
\end{gather}
where $n \in \mathbb{N}$ and $D =\left [x -\frac{1}{a},
x+\frac{1}{a} \right]$. In this case, the $n-$moments are
\begin{equation}
R_{m} = \frac{1-(-1)^{m}}{2} \frac{r_{m}}{a^{m}}.
\end{equation}
For any even $n$ one gets the system~(\ref{phase_even}). If $n$ is
odd, setting $\alpha = 1+ 2 \nu$ we obtain the system
\begin{gather}
\label{phase_odd}
\begin{aligned}
S_{x}^{2} &= N_{0} \\
S_{z} &= \frac{i \; r_{1}}{2} N_{0} S_{x} + \frac{1}{2} N_{1}.
\end{aligned}
\end{gather}
Compatibility condition is equivalent to the equation
\begin{equation}
N_{0z} = i\frac{3}{2} r_{1} N_{0} N_{0x} \pm \sqrt{N_{0}} N_{1 x}.
\end{equation}

\section*{Acknowledgments}
A.M. is pleased to thank Prof. W. Kr\'olikowski for useful
discussions.

\appendix

\end{document}